\documentclass[final,number,sort&compress,times]{elsart}

\usepackage{graphicx}

\usepackage{amssymb}
\usepackage{float}

\journal{Physics Letters A}

\begin{document}

\begin{frontmatter}

\title{Arbitrarily large numbers of kink internal modes in inhomogeneous sine-Gordon equations}

\author{J. A. Gonz\'alez}
\ead{jalbertgonz@yahoo.es}
\address{Department of Physics, Florida International University,\\ Miami, Florida 33199, USA and Department of Natural Sciences, Miami Dade College, 627 SW 27th Ave., Miami, Florida 33135, USA}
\author{A. Bellor\'{\i}n}
\ead{alberto.bellorin@ucv.ve}
\address{Escuela de F\'{\i}sica, Facultad de Ciencias, Universidad Central de Venezuela,\\ Apartado Postal 47586, Caracas 1041-A, Venezuela}
\author{M. A. Garc\'{\i}a-\~{N}ustes}
\ead{monica.garcia@pucv.cl}
\address{Instituto de F\'{\i}sica, Pontificia Universidad Cat\'olica de Valpara\'{\i}so, Casilla 4059, Chile}
\author{L. E. Guerrero}
\ead{lguerre@usb.ve}
\address{Departamento de F\'{\i}sica, Universidad Sim\'on Bol\'{\i}var,\\ Apartado Postal 89000, Caracas 1080-A, Venezuela}
\author{S. Jim\'enez}
\ead{s.jimenez@upm.es}
\address{Departamento de Matem\'atica Aplicada a las TT.II., E.T.S.I. Telecomunicaci\'on,\\ Universidad Polit\'ecnica de Madrid, 28040-Madrid, Spain}
\author{L. V\'azquez}
\ead{lvazquez@fdi.ucm.es}
\address{Departamento de Matem\'atica Aplicada, Facultad de Inform\'atica,\\ Universidad Complutense de Madrid, 28040-Madrid, Spain}

\begin{abstract}
We prove analytically the existence of an infinite number of internal (shape) modes of
sine-Gordon solitons in the presence of some inhomogeneous long-range
forces, provided some conditions are satisfied.
\end{abstract}


\end{frontmatter}

\section{\label{sec:Introduction}Introduction}

Solitary waves play a crucial role in many physical phenomena \cite%
{Buijnsters,Kim,Backes,Costa,Mendonza,Braun,Bishop,Kivshar,McLaughlin,Peyrard,Campbell,Campbell2,Kivshar2,Kivshar3,Rice,Boesch,Tchofo,Zhang,Zhang2,Zhang3,Majernikova,Quintero,Gonzalez,Kalbermann,Barashenkov,Oxtoby,Cuenda,Gonzalez2,Gonzalez3,Gonzalez4,Gonzalez5,Gonzalez6,Garcia,Gulevich,Gumerov,Dong,Zhang4,Liang,Cai,Liang2,Cai2,Gumerov2,Ekomasov,Morales-Molina,Quintero2,Gonzalez7,Chacon,Goldberg,Jiang,Sagues,Wang,Li,Abdullaev,Valenti,deBrito,Mendonza2,Bernardini}%
. Recently there has been much discussion about the existence of internal
(shape) modes of solitary waves \cite%
{Peyrard,Campbell,Campbell2,Kivshar2,Kivshar3,Rice,Boesch,Tchofo,Zhang,Zhang2,Zhang3,Majernikova,Quintero,Gonzalez,Kalbermann,Barashenkov,Oxtoby,Cuenda,Gonzalez2,Gonzalez3,Gonzalez4,Gonzalez5,Gonzalez6,Garcia,Gulevich,Gumerov,Dong,Zhang4,Liang,Cai,Liang2,Cai2,Gumerov2,Ekomasov,Morales-Molina,Quintero2}%
, and several novel physical effects have been discovered in nonintegrable
models possessing internal (shape) modes \cite%
{Gonzalez2,Gonzalez3,Gonzalez4,Gonzalez5,Gonzalez6,Garcia,Gulevich,Gumerov,Dong,Zhang4,Liang,Cai,Liang2,Cai2,Gumerov2,Ekomasov,Morales-Molina,Quintero2,Gonzalez7,Chacon,Goldberg,Jiang,Sagues,Wang,Li,Abdullaev,Valenti,deBrito,Mendonza2,Bernardini}%
.

The sine-Gordon equation, which is an integrable partial differential
equation that does not have internal (shape) modes, is one of the paradigms
in soliton theory. This is a fundamental model in many areas of physics. For
instance, it describes dislocations in crystals and fluxons in long
Josephson junctions, just to mention two examples \cite%
{Braun,Bishop,Kivshar,McLaughlin}.

Integrable models describe real physical systems only with certain
approximation \cite{Kivshar3}. So it is very important to study the
sine-Gordon equation in the presence of perturbations. Many papers have been
dedicated to the study of periodic time-dependent perturbations in the
sine-Gordon model \cite%
{Rice,Boesch,Tchofo,Zhang,Zhang2,Zhang3,Majernikova,Quintero}. The possible
observation of a \textquotedblleft quasimode\textquotedblright\ in this
model has been debated intensively in literature \cite%
{Rice,Boesch,Tchofo,Zhang,Zhang2,Zhang3,Majernikova,Quintero,Gonzalez,Kalbermann,Barashenkov,Oxtoby,Cuenda}%
. Some of these works \cite{Quintero,Cuenda} affirm that the mentioned
\textquotedblleft quasimode\textquotedblright\ is a numerical artifact.

In Ref. \cite{Gonzalez}, the authors have shown the existence of an internal
(shape) mode of the sine-Gordon kink when it is in the presence of
inhomogeneous space-dependent external forces, provided that these forces
possess some particular properties.

In the present letter we will investigate the inhomogeneous sine-Gordon
equation 
\begin{equation}
\phi _{tt}-\phi _{xx}+\sin \phi =F(x),  \label{Eq:1}
\end{equation}%
where the external perturbation $F(x)$ decays as a power-law.

We will prove the existence of an infinite number of shape modes of
sine-Gordon kinks in the presence of some long-range forces.

The paper is organized as follows. In Section \ref{sec:internalmodes}, we
introduce some important concepts for the understanding of the paper. In
Section \ref{sec:Exponentially-decaying}, we discuss the sine-Gordon
equation perturbed by an exponentially-decaying force. In Section \ref%
{sec:power-law}, we analyze the behavior of the kink in the presence of
perturbations with a power-law decay. Finally, in Section \ref%
{sec:Conclusions}, we present the conclusions of our analysis.

\section{\label{sec:internalmodes}Internal (shape) modes}

Before discussing the new results, we would like to clarify some important
concepts.

Consider the general equation 
\begin{equation}
\phi _{tt}-\phi _{xx}+G(\phi )=F(x).  \label{Eq:1*}
\end{equation}

Suppose this equation supports a kink solution $\phi _{k}(x)$ \cite%
{Peyrard,Campbell,Campbell2,Kivshar3,Rice,Gonzalez2,Goldstone,Currie,Jeyadev,Fogel1,Fogel2}%
.

To investigate the stability of the kink \cite%
{Peyrard,Campbell,Campbell2,Kivshar3,Rice,Gonzalez2,Goldstone,Currie,Jeyadev,Fogel1,Fogel2}%
, we look for solutions in the form $\phi (x,t)=\phi _{k}(x)+f(x)e^{\lambda
t}$. The perturbation $f(x)$ must satisfy a Schr\"{o}dinger-like equation 
\begin{equation}
\left[ -\partial _{xx}+U(x)\right] f(x)=\Gamma f(x),  \label{Eq:2*}
\end{equation}%
where $U(x)=\partial G\left( \phi _{k}(x)\right) /\partial \phi $, $\Gamma
=-\lambda ^{2}$.

In general, Eq. (\ref{Eq:2*}) has a discrete spectrum and a continuous
spectrum \cite%
{Peyrard,Campbell,Campbell2,Kivshar3,Rice,Gonzalez2,Goldstone,Currie,Jeyadev,Fogel1,Fogel2}%
.

The soliton modes are described by the eigenfunctions corresponding to the
discrete spectrum \cite{Gonzalez5}. On the other hand, the eigenfunctions
corresponding to the continuous spectrum are generally called the phonon
modes \cite%
{Peyrard,Campbell,Campbell2,Kivshar3,Rice,Gonzalez2,Goldstone,Currie,Jeyadev,Fogel1,Fogel2}%
. So the bound states of Eq. (\ref{Eq:2*}) are the soliton modes.

The soliton modes and the phonon modes determine the form of the
oscillations around the kink solution.

Let us discuss the $\phi ^{4}$ equation as an example: 
\begin{equation}
\phi _{tt}-\phi _{xx}-\phi +\phi ^{3}=0.  \label{Eq:3*}
\end{equation}

The static kink solution is $\phi _{k}(x)=\tanh (x/\sqrt{2})$.

The equation for $f(x)$ is 
\begin{equation}
-\partial _{xx}f(x)+\left[ 2-3\mathrm{sech}^{2}(x/\sqrt{2})\right] f(x)=\Gamma
f(x).  \label{Eq:4*}
\end{equation}

The eigenvalues and eigenfunctions are given by the well-known expressions:
for $\Gamma _{0}=0$ we get $f_{0}\left( x\right) =\mathrm{sech}^{2}(x/\sqrt{2}%
) $, for $\Gamma _{1}=3/2$ we obtain $f_{1}\left( x\right) =\tanh (x/\sqrt{2}%
)\mathrm{sech}(x/\sqrt{2})$, and $\Gamma _{k}=2+k^{2}$ correspond to the
continuum spectrum. These are the (unnormalized) wave functions.

These solutions correspond to the translation mode ($f_{0}\left( x\right) $%
), the internal (shape) mode ($f_{1}\left( x\right) $), and the continuum
phonons scattered by the kink \cite%
{Peyrard,Campbell,Campbell2,Kivshar3,Rice,Gonzalez2,Goldstone,Currie,Jeyadev,Fogel1,Fogel2}%
. In this case, the translation mode and the internal (shape) mode are the
soliton modes.

The internal (shape) mode is responsible mostly for the vibrations of the
kink width. The internal (shape) mode, with frequency $\omega =\sqrt{\Gamma }%
=\sqrt{3/2}$, represents a localized deformation around the kink and
describes oscillations of the kink shape. This mode has been used to explain
the complex resonance phenomena that occur during kink-antikink collisions 
\cite%
{Peyrard,Campbell,Campbell2,Kivshar3,Rice,Gonzalez2,Goldstone,Currie,Jeyadev,Fogel1,Fogel2}%
. We should remark that there are systems with several internal (shape)
modes \cite{Campbell,Gonzalez,Gonzalez2,Gonzalez8}.

The continuum corresponds physically to dispersive travelling waves that
propagate to spatial infinity \cite{Campbell}.

Another very important example is the pure sine-Gordon equation 
\begin{equation}
\phi _{tt}-\phi _{xx}+\sin \phi =0.  \label{Eq:5*}
\end{equation}

Various solutions are known \cite%
{Peyrard,Campbell,Campbell2,Kivshar3,Rice,Gonzalez2,Goldstone,Currie,Jeyadev,Fogel1,Fogel2}%
. For instance, the static kink is 
\begin{equation}
\phi _{k}(x)=4\arctan \left[ \exp \left( x\right) \right] .  \label{Eq:6*}
\end{equation}

In order to study the oscillation modes around the kink, we look for
solutions in the form $\phi (x,t)=\phi _{k}(x)+f(x)e^{\lambda t}$, where $%
f(x)$ satisfies the equation 
\begin{equation}
-\partial _{xx}f(x)+\left( 1-2\mathrm{sech}^{2}(x)\right) f(x)=\Gamma f(x).
\label{Eq:7*}
\end{equation}

Eq. (\ref{Eq:7*}) has exactly only one bound eigenstate with $\Gamma =0$ and 
\begin{equation}
f_{0}(x)=\mathrm{sech}(x).  \label{Eq:8*}
\end{equation}

The zero-frequency bound state is the Goldstone \cite{Goldstone} mode (also
called the translation mode). The remaining eigenfunctions form a continuum
with $\Gamma _{k}=1+k^{2}$. There are no internal (shape) modes in this
case, there is only one soliton mode.

The birth of new internal (shape) modes have been discussed in several
papers \cite{Kivshar3,Gonzalez,Gonzalez2,Gonzalez8}. An appropriate
perturbation can create a soliton internal (shape) mode \cite%
{Peyrard,Campbell,Campbell2,Kivshar3,Rice,Gonzalez2,Goldstone,Currie,Jeyadev,Fogel1,Fogel2}%
.

The existence of several internal (shape) modes in the perturbed $\phi ^{4}$
equation is studied in Ref. \cite{Gonzalez2} and the creation of several
internal (shape) modes in the perturbed sine-Gordon equation is studied in
Refs. \cite{Gonzalez,Gonzalez8}. The birth of an internal (shape) mode in
the double sine-Gordon equation is presented in Ref. \cite{Kivshar3}.

These phenomena can occur only when a new eigenvalue of the discrete
spectrum is created \cite{Kivshar3}.

\section{\label{sec:Exponentially-decaying}Exponentially-decaying
perturbation}

The model 
\begin{equation}
\phi _{tt}-\phi _{xx}+\sin \phi =F_{1}(x),  \label{Eq:2}
\end{equation}%
where $F_{1}(x)=2(B^{2}-1)\sinh (Bx)/\cosh ^{2}(Bx)$, was introduced in Ref. 
\cite{Gonzalez}. We recall here briefly the results presented in that work.

This force creates an equilibrium position for the center of mass of the
kink. The equilibrated kink solution is $\phi _{k}=4\arctan \left[ \exp
\left( Bx\right) \right] $. The stability investigation of this solution $%
\left[ \phi (x,t)=\phi _{k}(x)+f(x)e^{\lambda t}\right] $ can be expressed
as the eigenvalue problem $\widehat{L}f=\Gamma f$, where $\widehat{L}%
=-\partial _{x}^{2}+\left( 1-2\cosh ^{-2}(Bx)\right) $ and $\Gamma =-\lambda
^{2}$. This eigenvalue problem can be solved exactly \cite%
{Gonzalez,Gonzalez2,Flugge}. The eigenvalues of the discrete spectrum can be
calculated%
\begin{equation}
\Gamma _{n}=B^{2}\left( \Lambda +2\Lambda n-n^{2}\right) -1,  \label{Eq:3}
\end{equation}%
where $\Lambda =-\frac{1}{2}+\sqrt{\frac{1}{4}+\frac{2}{B^{2}}}$ \cite%
{Gonzalez,Gonzalez2,Flugge}. The smallest value of $n$ is zero. The largest
value of $n$ is $n_{\max }=\left[ \Lambda \right] -1$, where $\left[ \Lambda %
\right] $ is the integer part of $\Lambda $. The number of discrete
eigenvalues is equal to the integer part of $\Lambda $.

If $1/3<B^{2}\leq 1$, the integer part of $\Lambda $ is one ($\left[ \Lambda %
\right] =1$). So there is only one soliton mode. In the case $B^{2}=1$, the
Eq. \ref{Eq:2} possesses translational invariance, and this state is called
translational mode.

If $1/6<B^{2}\leq 1/3$, then $\left[ \Lambda \right] =2$. So a new internal
(shape) mode is created. If we continue decreasing parameter $B$, new
eigenvalues of the discrete spectrum are created. All the new modes are
internal (shape) modes.

\section{\label{sec:power-law}Perturbation with a power-law decay}

Consider now the completely new model 
\begin{equation}
\phi _{tt}-\phi _{xx}+\sin \phi =F_{2}(x),  \label{Eq:4}
\end{equation}
where $F_{2}(x)=2B\left[ \left( 2B^{2}-1\right) x-B^{2}x^{3}\right] /\left(
1+B^{2}x^{2}\right) ^{2}$.

Depending on parameter $B$, the perturbation $F_{2}(x)$ can have one or
three zeroes.

For $(2B^{2}-1)>0$, the function $F_{2}(x)$ has three zeroes. Otherwise, the
only zero is $x=0$.

We can find an exact solution for the stationary kink whose center of mass
is at point $x=0$: 
\begin{equation}
\phi _{k}=2\arctan (Bx)+\pi .  \label{Eq:5}
\end{equation}

As $\partial \phi _{k}/\partial x=2B/\left[ 1+\left( Bx\right) ^{2}\right] $%
, it is easy to see that for large values of $\left\vert x\right\vert $ the
solution behaves as a power law. For instance, as $x\rightarrow \infty $, $%
\left( \phi _{k}-2\pi \right) \sim 1/x$.

Compare this with the kink solution of Eq. (\ref{Eq:2}) that possesses an
exponential asymptotic behavior $\left( \phi _{k}-2\pi \right) \sim e^{-Bx}$%
. Note that parameter $B$ determines how fast the solution approaches the
value $2\pi $ for large values of $x$. On the other hand, for $B\rightarrow
0 $, the integer part of $\Lambda $ ($\left[ \Lambda \right] $) (which gives
the number of soliton modes) behaves as $\left[ \Lambda \right] \sim 1/B$.
Thus the asymptotic behavior of the kink solution is related to the number
of internal (shape) modes.

The stability analysis of solution (\ref{Eq:5}) leads to the following
eigenvalue problem 
\begin{equation}
\widehat{L}f=\Gamma f,  \label{Eq:6}
\end{equation}
where $\widehat{L}=-\partial _{xx}+W(x)$, $W(x)=1-2/\left(
1+B^{2}x^{2}\right) $.

Note that this is a Schr\"{o}dinger-like equation (compare Eq. (\ref{Eq:6})
with the one-dimensional time-independent Schr\"{o}dinger equation, $-\frac{%
\hbar^{2}}{2m}\frac{\partial ^{2}\psi }{\partial x^{2}}+V(x)\psi =E\psi $). As
discussed above, the bound states correspond to the soliton modes.

Let us re-write the Eq. (\ref{Eq:6}) in the following form 
\begin{equation}
\widehat{M}f=\Delta f,  \label{Eq:7}
\end{equation}
where $\widehat{M}=-\partial _{xx}+V(x)$, $V(x)=-2/(1+B^{2}x^{2})$, and $%
\Delta =\Gamma -1$.

This will allow us to apply directly some important theorems \cite{Chadan}.

Using results from Ref. \cite{Chadan} we will reformulate here a theorem in
a way that will be useful for our study of Eq. (\ref{Eq:7}).

Define the Schr\"{o}dinger equation $-\partial _{xx}f+V(x)f=\Delta f$ and
suppose that $V(x)$ is bounded from below, and that $V(x)\rightarrow 0$ as $%
\left\vert x\right\vert \rightarrow \infty $.

The number of bound states is infinite if there exists an $x_{\ast }>0$ such
that 
\begin{equation}
x^{2}V(x)<-1/4,  \label{Eq:8}
\end{equation}
both for $x>x_{\ast }$ and for $x<-x_{\ast }$.

On the other hand, if there is an $x_{\ast }>0$ such that $x^{2}V(x)>-1/4$
for $\left\vert x\right\vert >\left\vert x_{\ast }\right\vert $, then, the
number of bound states is finite.

Let us apply this theorem to our Eq. (\ref{Eq:7}).

We choose $x_{\ast }=\sqrt{1/(8-B^{2})}$, where $B^{2}<8$.

As $-2x^{2}/(1+B^{2}x^{2})<-1/4$ for $x>x_{\ast }$ and for $x<-x_{\ast }$,
then we have proved that we can find an $x_{\ast }$\ such that $%
V(x)=-2/(1+B^{2}x^{2})$ satisfies the conditions (\ref{Eq:8}).

For $B^{2}>8$, the number of bound states is always finite.

The remarkable result is that if $B^{2}<8$, then the perturbed sine-Gordon
equation (\ref{Eq:4}) can possess an arbitrarily large number of internal
(shape) modes!

\section{\label{sec:Conclusions}Conclusions}

We have investigated the inhomogeneous sine-Gordon equation given by Eq. (%
\ref{Eq:1}).

We have found exact kink solutions to the perturbed sine-Gordon equation,
and we have been able to study analytically the kink stability problem.

Our conclusion is that a kink equilibrated by an exponentially-localized
perturbation has a finite number of oscillation modes, whereas a
sufficiently broad equilibrating perturbation supports an infinite number of
soliton internal (shape) modes. This phenomenon is particularly relevant in
systems with inhomogeneous long-range forces.

We believe that the soliton-like structures described in this paper, which
essentially are extended objects with internal (shape) mode oscillations,
can play an important role in both Particle Physics and Condensed Matter
Theory.

M. A. Garc\'{\i}a-\~{N}ustes thanks the financial support of FONDECYT
project 11130450.

\bibliographystyle{model3-num-names}

\end{document}